\documentclass[reprint,twocolumn,amsmath,amssymb,aps,
]{revtex4}

\usepackage{epsfig,bm,epsf,graphics}
\usepackage{graphicx}
\usepackage{dcolumn}
\usepackage{bm}

\begin{document}
\preprint{APS/123-QED}

\title{Magnetic Properties of (La$_{0.56}$Ce$_{0.14}$)Sr$_{0.30}$MnO$_{3}$ Perovskite}
\author{Samia Yahyaoui\footnote{yahyaoui.samia@yahoo.fr}}
\affiliation{%
Unit\'e de Recherche de Physique Quantique, D\'epartement de Physique,
Facult\'e des sciences de Monastir, BP 22, 5019 Monastir, Tunisia.\\}
\author{H. T. Diep\footnote{diep@u-cergy.fr}}
\affiliation{%
Laboratoire de Physique Th\'eorique et Mod\'elisation,
Universit\'e de Cergy-Pontoise, CNRS, UMR 8089\\
2, Avenue Adolphe Chauvin, 95302 Cergy-Pontoise Cedex, France.\\
}%

\date{\today}

\begin{abstract}
We investigate in this paper magnetic properties of the perovskite compound (La$_{0.56}$Ce$_{0.14}$)Sr$_{0.30}$MnO$_{3}$.
The method we use here is Monte Carlo simulation, in which we take into account different kinds of interactions between nearest and between next-nearest
magnetic ions  Mn$^{3+}$ ($S$ = 2) ,  Mn$^{4+}$ ($S$ = 3/2) and  Ce$^{3+}$ ($S$ = 1/2). Using a classical spin model,
we have calculated the internal energy, the magnetization per ion type and their corresponding magnetic susceptibility, as well as the
Edwards-Anderson order parameter for each ion kind. We also studied the applied-field effect on the system magnetization.
Our results show a good agreement with experiments.
\vspace{0.5cm}
\begin{description}
\item[PACS numbers:75.30.-m , 75.50.-y , 75.10.Hk , 05.10.Ln]
\end{description}
\end{abstract}

\pacs{PACS numbers: XXX}
\maketitle

\section{INTRODUCTION}

\indent Many efforts have been recently devoted to investigate magnetic semiconductors
\cite{Dietl,Dietl2000,Grahn} and insulators
which could be used in spintronics device applications \cite{Wolf,Wu,Dagotto,Stanley,Magnin}.
Magnetic properties of such systems have been extensively studied, theoretically, experimentally and numerically \cite{Mac,Zhang}.
Perovskite-type manganese oxides \cite{Martin,Coey} have attracted much attention because of their interesting physical properties such
as colossal magnetoresistance \cite{Dagotto,Fert,Grunberg} and orbital ordering \cite{Kusters}. The undoped perovskite manganite LaMnO$_{3}$
is a charge-transfer type insulator \cite{Zaanen} which shows an A-type
antiferromagnet. When a percentage of La atoms are replaced by Sr atoms, the compound has a rich electronic phase diagram, including a doping-dependent metal-insulator transition, paramagnetism and ferromagnetism \cite{Urushibara}. La$_{1-x}$Sr$_x$MnO3 is one of the perovskite manganites that shows the colossal magnetoresistance effect \cite{Ramirez} and is also an observed half-metal for compositions around $x$=0.3 \cite{Park}.
In the manganite LaSrMnO$_3$ the ratio La$^{3+}$/Sr$^{2+}$ determines the oxidation state of Mn
and thus the ratio Mn$^{3+}$/Mn$^{4+}$. This corresponds to the number of Mn sites with a single
orbital occupied.  Depending on the composition, different magnetic ground states are observed \cite{Martin,Coey}.

The observed magnetic and electrical concomitant transitions can be qualitatively
understood by invoking double exchange (DE) mechanisms. The nature of the magnetic ordering in the entire compositional range depends on the relative concentrations of Mn$^{3+}$
and Mn$^{4+}$ and also on the structural properties.
According to the DE mechanism, the spin hopping from  Mn$^{3+}$ to Mn$^{4+}$ via the O$^{2-}$
orbital and leads to an effective ferromagnetic interaction.  Experiments have determined many magnetic properties. However, these
 experimental works allow only a partial understanding of the compound.
A better understanding of the roles of each microscopic interaction in
of this magnetic system requires more theoretical and numerical studies.
Let us mention a few works in this direction. The one- and two-orbital DE models for manganites have been studied using Monte Carlo (MC)
techniques in the presence of a robust electron-phonon
coupling (but neglecting antiferromagnetic exchange interactions between the localized spins) \cite{Sen1}.
A recent result \cite{Sen2,Trukhanov} shows that super-exchange (SE) interaction is
indispensable to provide an elastic model for manganites.

In this paper, we shall use MC simulations to study the compound. Note that the oxygen is coupled to full outer shell states Mn$^{3+}$ and Mn$^{4+}$, namely this a mixed valence
compound:  A$^{3+}_{1-x}$B$^{2+}_{x}$Mn$^{3+}_{1-x}$Mn$^{4+}_{x}$O$_{3}$.
In order to study the effect of substitution by cerium at the A-site because cerium
can exist in tri-, tetra and mixed-valence states. In this work we shall confirm as seen below that
Ce ions exist in a trivalent state in Sr-containing manganites,
in agreement with results reported in the literature \cite{Sundaresen1,Sundaresen2,Kallel}.
Using MC simulations we have determined magnetic properties of perovskite
manganite (La$_{1-x}$Ce$_{x}$)$_{0.70}$Sr$_{0.30}$MnO$_{3}$, taking $x$ = 0.20.
We carry out numerical simulations using an Ising-like spin model with various interactions based on experiment observations. This model is justified
by an excellent agreement with experimental measurements performed on this material. \\
\indent In Sec. 2, we present our model and describe our MC method.
Results are reported and discussed in Sec. 3. Concluding remarks are given in
Sec . 4.

\section {Model and Method}
\subsection {Model}\label{model}

\indent To describe this system, we use the body-centered-cubic (BCC) lattice with the following  Hamiltonian:
\begin{equation}
 {\cal H} = -\sum_{<i,j>}J_{ij} \mathbf {S}_{i}\cdot \mathbf {S}_{j}-\mu_{0} \sum_{<i>}\mathbf H\cdot \mathbf {S}_i
\end{equation}
where $\mathbf S_i$ is the spin at the lattice site i, $\sum_{<i,j>}$ is made over spin pairs coupled through
the exchange interaction $J_{ij}$ with $J_{ij} <0$ for antiferromagnetic interactions and
$J_{ij} >0$ for ferromagnetic interactions.
In the following, we shall take interactions between nearest-neighbors (NN) and between next-nearest
neighbors (NNN) of magnetic ions. $\mathbf H$ is a magnetic field applied along the $z$ axis.
According to results given by N. Kallel {\it et al.} \cite{Kallel}, the spin
magnitudes of Mn$^{3+}$, Mn$^{4+}$ and Ce$^{3+}$ are $S = 2$, $S = 3/2$ and $S = 1/2$, respectively.
The DE mechanism between Mn$^{3+}$-O-Mn$^{4+}$ has been recognized \cite{Zener1} although some debate continues.
This characteristic is related with a new interesting observed ferromagnetic
transition in doped manganites \cite{Hong,Bose}. As far as the doped sample
(La$_{0.56}$Ce$_{0.14}$)Sr$_{0.30}$MnO$_{3}$ is concerned, the magnetization investigation
presents a very sharp ferromagnetic-paramagnetic transition at 357 K. \\

Before defining explicitly the interactions, let us define the spin model.
Using an Ising-like spin model we obtain  an excellent agreement with experiments.
The exchange parameters $J_{ij}$ are strongly correlated to the electronic structure of the compound.
In this materials we consider that the ferromagnetic
 double exchange (Mn$^{3+}$-Mn$^{4+}$ ) and antiferromagnetic super-exchange interactions
 (Mn$^{3+}$-Mn$^{3+}$, Mn$^{4+}$-Mn$^{4+}$, Mn$^{3+}$-Ce$^{3+}$
and Mn$^{4+}$-Ce$^{3+}$ ) between transition metals are competed \cite{Goodenough,Kanamori,Eto,Zener2,Zhou}. On the other hand,
the interaction between Ce$^{3+}$-Ce$^{3+}$ is considered as ferromagnetic.
Note that SE interactions Mn-Ce was not explained by Kanamori and Goodenough but it has been explained by other researchers \cite{Sundaresen1,Sundaresen2,Yahyaoui2015}.

Based on the crystal and electronic structure of this system,
the NN interactions can be taken  into account in the present study are:\\

\indent $J_{1}$: Interaction Mn$^{3+}$-Mn$^{4+}$.\\
\indent $J_{2}$: Interaction Mn$^{3+}$-Mn$^{3+}$.\\
\indent $J_{3}$: Interaction Mn$^{4+}$-Mn$^{4+}$.\\
\indent $J_{4}$: Interaction Mn$^{3+}$-Ce$^{3+}$.\\
\indent $J_{5}$: Interaction Mn$^{4+}$-Ce$^{3+}$.\\
\indent $J_{6}$: Interaction Ce$^{3+}$-Ce$^{3+}$.\\

In addition, we also introduce the following NNN interactions:\\

\indent $J_{7}$: NNN interaction Mn$^{3+}$-Mn$^{3+}$.\\
\indent $J_{8}$: NNN interaction Mn$^{3+}$-Mn$^{4+}$.\\
\indent $J_{9}$: NNN interaction Mn$^{4+}$-Mn$^{4+}$.\\

Experimental observations suggested that $J_1$ is ferromagnetic and much larger than $J_2$ and $J_3$ which are antiferromagnetic \cite{Kallel}.  It was also suggested that $J_4$ and $J_5$ are very small and antiferromagnetic while $J_6$ is very small but ferromagnetic.  These suggestions will help to retain only essential interactions in the following.  Note that the NNN interactions $J_7$, $J_8$ and $J_9$, though also very small, play an important role in the low-$T$ behavior of the compound.   We note that the values of the exchange integrals given above will be deduced from experimental data by fitting the MC transition temperature as seen below.

In order to explain the decrease of the global magnetization at low temperatures observed in the compound \cite{Kallel}, we introduce into the system a small number of cluster of 7 to 10 spins.
This is justified by the fact that in doped systems the presence of such clusters cannot be avoided whatever doping methods are \cite{Auslender,Baldini,Kovaleva,Allieta}). It should be noted that a cluster is composed of spins strongly connected to each other by a strong interaction with respect to interactions between its outer spins with the host matrix. At high temperatures these clusters are free to flip because the bonds connecting the clusters to the host matrix are broken by the temperature. So the cluster magnetization does not affect the global magnetization of the compound at high $T$. However, at low-enough temperatures, these bonds resist to $T$ and as a consequence the clusters are frozen in a direction. If the intra-cluster interaction is ferromagnetic and its outer bonds with the matrix spins are antiferromagnetic, then the clusters are frozen with its spins antiparallel to the spins of the host matrix. The global magnetization is therefore significantly reduced. That is what we observed in simulations as shown below. Note that the spins of Ce$^{3+}$ alone cannot make such a strong magnetization reduction not only because  their spin magnitude is small ($S=1/2$), but also because of their small number (20\% of the centered sites of the BCC lattice).

\subsection{Method}

\indent Simulations have been performed for systems of $N=2 L^3$ spins,
where $L$ is the number of BCC cells in each of the $x$, $y$ and $z$ directions.
We note that each lattice contains two types of Mn, namely Mn$^{3+}$ with spin $S=2$ and Mn$^{4+}$ with spin $S=3/2$,
and  Ce$^{3+}$ ions at the cube centers. We consider in this work their respective concentrations $x = 0.7$, 0.3 and 0.2,  since these are the ones which have been experimentally studied \cite{Kallel}.
We use the periodic boundary conditions in all three directions to avoid surface effects. To estimate finite-size effects, we use several different lattice sizes where
$L = 12$, $16$ and $20$. As will be discussed below, due to a strong disorder (random mixing of Mn ions), the size effects are not significant for $L>20$. Most of the simulation have been therefore carried out at this size with disorder average on many samples.
In order to determine various physical quantities as functions of temperature $T$,
we have performed a standard Metropolis MC simulation \cite{DiepTM,Metropolis,Binder}.
We are aware that more sophisticated MC methods such as histogram techniques \cite{Ferrenberg,Ferrenberg2,Pham} or Wang-Landau flat density simulations \cite{Wang-Landau,Ngo2008a,Ngo2008b}
could bring more relevant information, for instance more precise magnetic susceptibility obtained with larger system sizes. However, in this work we aimed at obtaining values of various exchange integrals and to see their effect. The rigorous Wang-Landau method, which is time-consuming, can be used in the next step.
The procedure of our simulation can be split into two steps. The first step consists in equilibrating
the lattice at given temperature. When equilibrium is reached,
we determine the thermodynamic properties by taking thermal averages of various physical quantities.
The MC run time for equilibrating is about $10^5$ MC steps per spin.
After equilibrating, the averaging is taken, over $10^5$ MC steps.
In MC simulations, when equilibrium
is reached, we have calculated the averaged total magnetic energy per spin $E$,
the magnetic susceptibility $\chi$, the magnetization of each ion type and the total magnetization,
as functions  of temperature $T$ and magnetic field $H$. These quantities are defined in the following:\\

The total energy \\
\begin{equation}
 {E} = {<{\cal H}>};\\
\end{equation}
The magnetization of ions of type $\ell$ \\
\begin{equation}
M_{\ell} = \frac{1}{N_{\ell}} <\sum_{i\in \ell}S_i >;\\
\end{equation}
The total magnetization $M_t$\\
\begin{equation}
M_t = \frac{1}{N} <\sum_{i}S_i>;\\
\end{equation}
The magnetic susceptibility per spin\\
\begin{equation}
\chi = \frac{N}{k_BT}[<{M_t^2}>-<M_t>^2];\\
\end{equation}
and the Edwards-Anderson order parameter $Q_{EA}(\ell)$ of the ion type $\ell$\\
\begin{equation}
Q_{EA}(\ell) = \frac{1}{N_\ell} \sum_{i\in \ell}|\sum_t S_i(t)|;\\
\end{equation}
where $<...>$ indicates the thermal average and the sum is taken over each ion type Mn$^{3+}$ ($\ell = 1$), Mn$^{3+}$ ($\ell = 2$) or
Ce$^{3+}$ ($\ell = 3$) with $N_{\ell}$ being the number of spins of each kind. Note that the
$Q_{EA}(\ell)$ is calculated by taking the time average of each spin before averaging over all spins of the subsystem.  This order parameter is very useful in the case of disordered systems such as spin glasses or doped compounds: it expresses the degree of freezing of spins independent of whether the system has a long-range order or not \cite{DiepTM}.

\section {Results}

We introduce
a doping at $20\%$ of Ce ions into the compound. Simulations have been performed and we used the critical temperature experimentally observed for this case $T_c=357$ K to estimate various exchange interactions listed above.  For that purpose, we use the mean-field formula \cite{DiepTM}
\begin{equation}\label{MFTC}
T_c=\frac{2}{3k_B}Z_{eff}S_{eff}(S_{eff}+1)J_{eff}
\end{equation}
where $Z_{eff}$ is the effective coordination number and $S_{eff}$  the effective spin value. $Z_{eff}$ is approximately taken by $Z_{eff}\simeq 6$ (Mn sites only, neglecting small number of Ce sites). $J_{eff}$ is on the other hand
calculated by
\begin{eqnarray}
S_{eff}&=&[0.3S(\mbox{Mn}^{4+})+0.7 S(\mbox{Mn}^{3+})\nonumber\\
&&+0.2S(\mbox{Ce}^{3+}]/(0.3+0.7+0.2)\simeq 1.61
\end{eqnarray}
where we have taken into account the concentrations of each species. Putting $T_c=357$ K in Eq. (\ref{MFTC}), we obtain $J_{eff}\simeq 21.24 $ K.  Note that in magnetic materials with Curie temperatures at room or higher temperatures the effective exchange interaction is of the order of several dozens of Kelvin \cite{Magnin,Restrepo-Parra,Pavlukhina}. In the works of Restrepo-Parra {\it et al.} and of Pavlukhina {\it et al.}, by fitting experimental values of the transition temperatures for different regions of doping concentration $x$,  the authors have deduced the values of different kinds of exchange interactions, in a similar manner to ours.  The model used in their work was a Heisenberg model with a uniaxial anisotropy, while our model is an Ising-like model.  Since their anisotropy is very strong (of the order of exchange interaction), their model is not far from ours: our results are in agreement with theirs provided the following changes in the Heisenberg spins and anisotropy value. At 30\%  Restrepo-Parra {\it et al.} used $J_1=5.18$ meV and 0.9 meV depending on the orbitals of Mn$^{3+}$-Mn$^{4+}$. The average is thus (5.18+0.9)/2=3.02 meV $\simeq$ 30 K. Our value of $J_1$ is $\simeq$ 21.24 K which is lower. However we note that the comparison is not rigorous because their material is LaCaMnO$_3$ which has $T_c$=258 K while ours is LaSrCeMnO$_3$ with $T_c=367$ K.
Anyway, the values of $J_1$ are found in the range of a few dozens of Kelvin for materials of high Curie temperature.

At this stage, note that it is not easy to determine each of the exchange interactions defined earlier. Fortunately, we know that $J_1$ is much larger than the other interactions \cite{Kallel}.
In the mean-field spirit, $J_{eff}$ is a linear combination of all interactions and taking, after comparing our calculated magnetizations in an applied field (see below) using several trying values we find the best fits are obtained with $J_2=J_3=-J_1/60$, $J_4=J_5=-J_1/12$ and $J_6=J_1/60$. The NNN interactions have been taken as $J_7=J_9=J_1/12$ and $J_8=J_1/60$.  The best estimated value of the main exchange interaction is $J_1 \simeq J_{eff}$  since contributions from other much smaller positive and negative interactions almost cancel out.

\begin{figure}[ht!]
\centering
\includegraphics[width=7cm,angle=0]{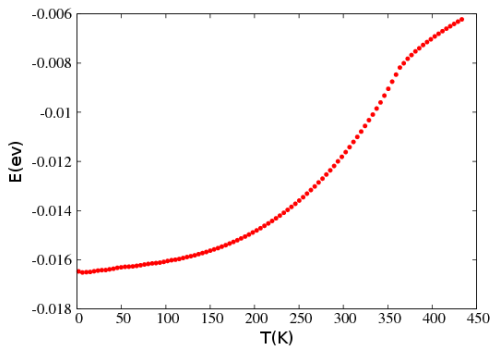}
\caption{(Color online) Top: energy $E$ (eV) versus $T$(K).  \label{figeq}}
\end{figure}

\begin{figure}[ht!]
\centering
\includegraphics[width=7cm,angle=0]{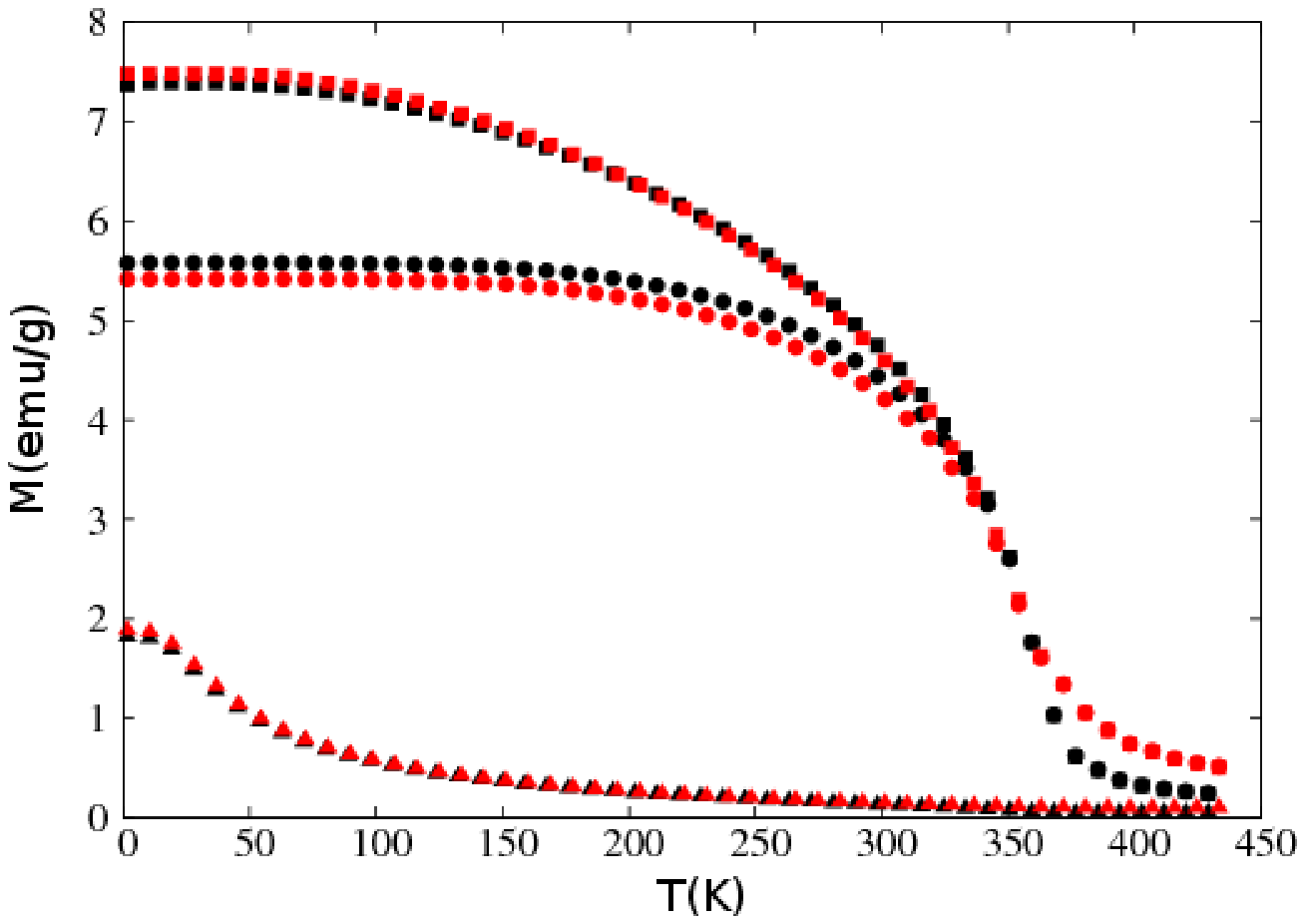}
\includegraphics[width=7cm,angle=0]{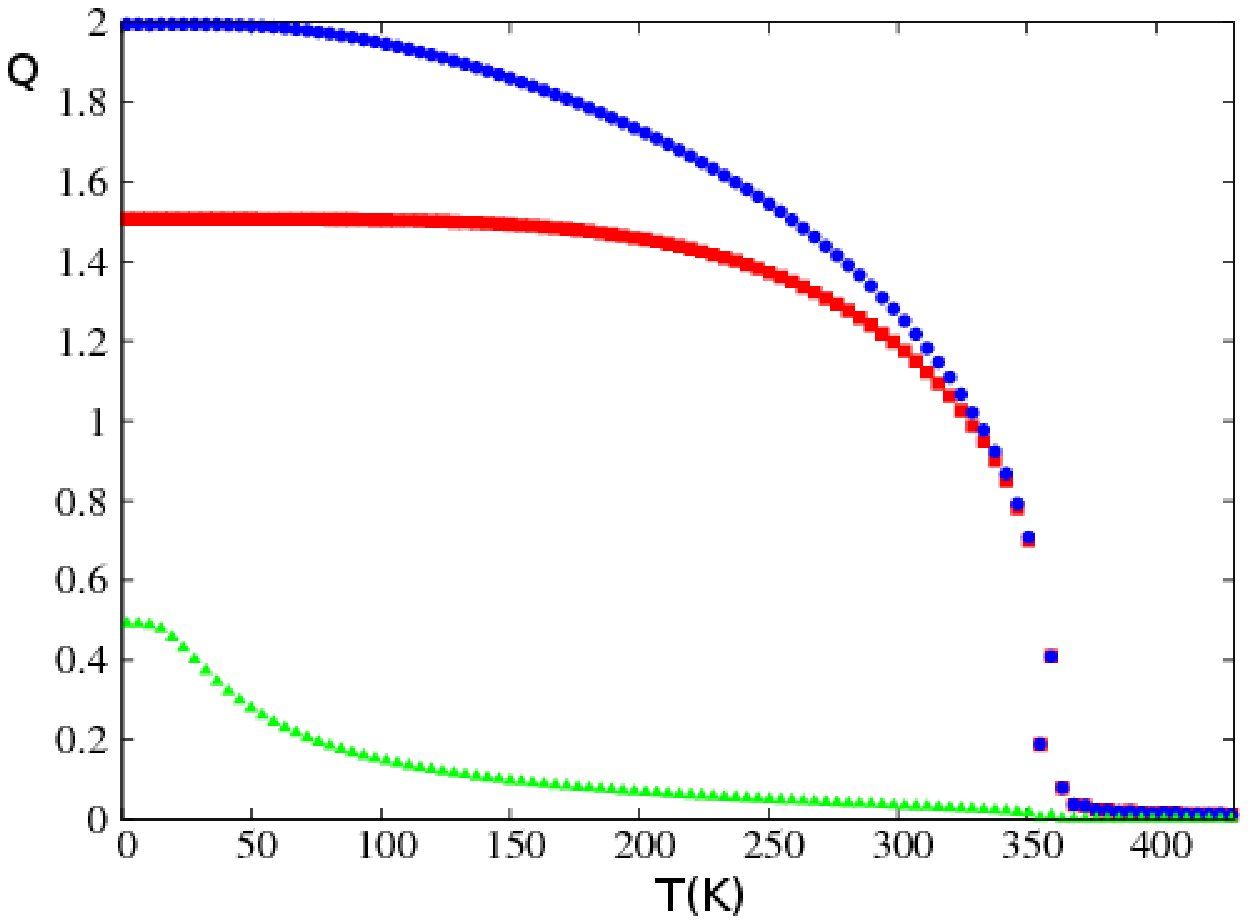}
\caption{(Color online) Top: Magnetizations of subsystems of Mn$^{3+}$ (squares), Mn$^{4+}$ (circles), and Ce$^{3+}$ (triangles) at two lattice sizes $N=2L^3$ with $L=12$ (red) and 20 (black).  Bottom: Edwards-Anderson order parameter $Q_{EA}$ of Mn$^{3+}$ ($S=2$, blue circles), Mn$^{4+}$ ($S=3/2$, red squares), and Ce$^{3+}$ ($S=1/2$, green triangles). See text for comments. \label{figm}}
\end{figure}

\begin{figure}[ht!]
\centering
\includegraphics[width=8cm,angle=0]{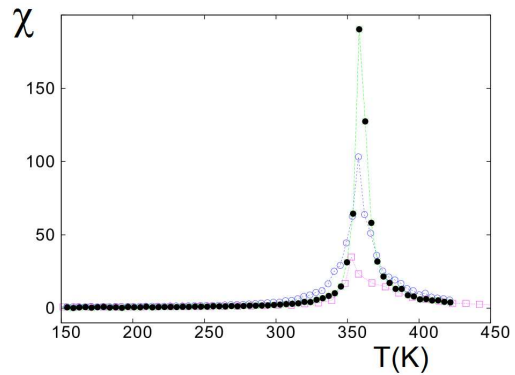}
\caption{(Color online) Magnetic susceptibility $\chi$ in unit of emu/g/Tesla versus temperature $T$ in Kelvin for three lattice sizes $L=12$ (magenta squares), 20  (blue void circles), 30 (black circles).  \label{figx}}
\end{figure}

The ground state magnetic energy per spin is written as $E_0=-\frac{1}{2} Z_{eff} J_{eff}S_{eff}^2$ \cite{DiepTM}. Using the values estimated above we obtain $E_0\simeq -0.0165$ eV.
The energy in unit of eV as a function of temperature is shown in Fig. \ref{figeq}.
One notes a strong change of
curvature at $T_c\simeq 357$ K signaling a phase transition occurring at this temperature.
This is confirmed by the magnetization curves and the order parameter $Q_{EA}$ shown in Fig. \ref{figm} as well as the peak of the magnetic susceptibility shown in
Fig. \ref{figx}. Note that in the latter figure,  results of three system sizes $L=12$, 20 and 30 are
displayed to show a strong finite-size effect between $L=12$ and 20. The peak becomes stronger for larger $L$, but the positions of $T_c$ do not significantly change between $L=20$  and 30.

Let us emphasize that since the exchange integrals $J_7$, $J_8$ and $J_9$ between NNN are very weak, these bonds are active only at low $T$.  The sensitivity of each of them is not the same. The $J_7$ between NNN Mn$^{3+}$ gives a most significant effect at $T<50$ K: due to the high Mn$^{3+}$ concentration (70\%),  Mn$^{3+}$ ions have a high probability to be surrounded by other Mn$^{3+}$ ions, making $J_7$ active.  Other ions (Mn$^{4+}$ and Ce$^{3+}$) are very diluted so that the number of NNN is fewer.   The effect of these NNN interactions is to make the ferromagnetic ordering stronger at low temperatures: curves of magnetizations of Mn$^{3+}$ and Mn$^{4+}$ shown in Fig. \ref{figm} are more "horizontal" for $T<150$ K as the experimental curves.  If we neglect the NNN interactions and use only the NN interactions,  the magnetization curves still fall down at $T_c=357$ K but their diminution with increasing temperature below $T_c$  is stronger with respect to the experimental curves.

\begin{figure}[ht!]
\centering
\includegraphics[width=7cm,angle=0]{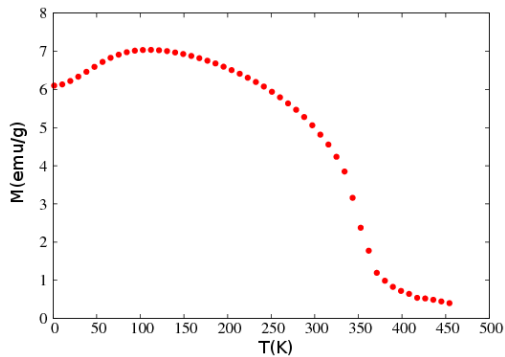}
\caption{(Color online) Total magnetizations versus $T$ with clusters embedded in the compound. System size: $N=2L^3$, number of 7-spin clusters: 2$L$, with $L=20$. See text for comments. \label{figmc}}
\end{figure}

In order to explain the strong decrease of the global magnetization at low temperatures, we introduce randomly a small number of clusters of size 7 spins, namely 1\% of the Mn sites. The total number of cluster spins is thus 7\% of the total system size.  Intra-cluster interaction is assumed to be ferromagnetic and equal to $J_1/10$ while its interaction with the host-matrix neighboring spins is antiferromagnetic and equal to  $-J_1/30$, independent of the type of ion, for simplicity.  The total magnetization is shown in Fig. \ref{figmc} where we observe a strong decrease from 7 emu/g to 6.1 emu/g in agreement with experiments in the temperature range around 100 K \cite{Kallel}. Note that at lower temperatures, experimental curve shows a sudden decrease that we cannot explain by the present model. Perhaps other kinds of clusters should be introduced to explain this behavior in the same spirit as the introduction of our clusters described above.  We note that as said at the end of subsect. \ref{model}, the introduction of clusters into the system does not alter the high temperature behavior, in particular the value of $T_c$.

\begin{figure}[ht!]
\centering
\includegraphics[width=7cm,angle=0]{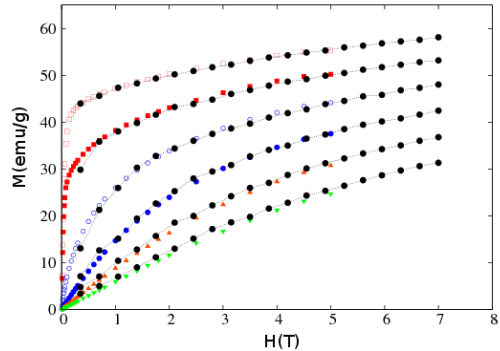}
\caption{(Color online) Monte Carlo results (black circles) of the total magnetization (in emu/g) versus applied field $H$ (in Tesla) are compared to experimental data available up to 5 Tesla at several temperatures shown from top: $T=340$ K (red void squares), 350 K (red solid squares), 360 K (blue void circles), 370 K (blue solid circles), 380 K (red triangles) and 390 K (green triangles).   An excellent agreement is observed at each $T$.\label{figh}}
\end{figure}

Let us show in Fig. \ref{figh} the response of the system to an applied magnetic field at several temperatures around $T_c$. We show in the same figure experimental data available at the same temperatures taken from Ref. \onlinecite{Kallel}. A remarkable agreement between our results and the experimental data is observed. This shows that the model used in this paper is very appropriate to describe the compound.

 \section {Conclusion}

\indent We have shown in this paper the MC results of the perovskite (La$_{0.56}$Ce$_{0.14}$)Sr$_{0.30}$MnO$_{3}$.  The model includes NN and NNN interactions between different types of magnetic ions in the compound. The main interaction which is responsible for the high Curie temperature is that between Mn$^{3+}$ and Mn$^{3+}$. Other interactions, though much smaller, help describe coherently the behavior of the system in several aspects.  We have introduced a random distribution of small clusters into the compound to explain low-$T$ behavior.
We obtained values of interactions by using only the experimental value of Curie temperature.
Besides, we have studied the applied-field effect on the magnetization in excellent agreement with experiments.  In view of this agreement, we conclude that the assumption used in our model on the trivalent state Ce$^{3+}$ is a correct one.\\

\acknowledgments
 SY acknowledges  a financial support from the Ministry of Education of Tunisia. She is grateful to the University of Cergy-Pontoise for hospitality where this work has been accomplished.

{}

\end{document}